\documentclass[preprint,5p,twocolumn]{elsarticle}
\usepackage{amssymb}
\usepackage{epsfig}
\usepackage{color}
\newcommand{\be}{\begin{equation}}
\newcommand{\ee}{\end{equation}}
\newcommand{\ba}{\begin{eqnarray}}
\newcommand{\ea}{\end{eqnarray}}

\usepackage{graphicx}
\usepackage{dcolumn}
\usepackage{bm}
\usepackage{hyperref}
\usepackage{hyperref}
\RequirePackage[normalem]{ulem} 
\RequirePackage{color}\definecolor{RED}{rgb}{1,0,0}\definecolor{BLUE}{rgb}{0,0,1} 

\begin{document}

\begin{frontmatter}



\title{Two-electron entanglement in elliptically deformed quantum dots}


\author{Przemys\l aw Ko\'scik and Anna Okopi\'nska
}
\address{Institute of Physics,  Jan Kochanowski University\\
ul. \'Swi\c{e}tokrzyska 15, 25-406 Kielce, Poland}

\begin{abstract}
Entropic entanglement measures of a two-dimensional system of two Coulombically interacting particles confined in an anisotropic harmonic potential are discussed in dependence on the anisotropy and the interaction strength. The harmonic approximation appears exact in the strong interaction limit, allowing determination of the asymptotic expression for the linear entropy. Entanglement properties are dramatically influenced by the anisotropy of the confining potential in the strong-correlation regime.

\end{abstract}


\end{frontmatter}


\section{Introduction}\label{1}
The Hookean system composed of Coulombically interacting particles confined in a harmonic potential is of increasing interest as it provides an effective model of semiconductor quantum dots (QDs)~\cite{fab}. Determination of the amount of entanglement in various states of such systems is important in view of their possible applications in quantum information technology~\cite{QCdots}. The simplest candidate for studying the entanglement properties is the two-particle Hookean system. Although the system was considered in various theoretical contexts
both in 2D \cite{hf,anise,anise3,anise4,kos,anise1,high-dens} and in 3D case \cite{high-dens,kas,Karw,ent2},
the influence of the confinement anisotropy on entanglement has not  been  investigated so far. In this paper we undertake the investigation of this issue, restricting ourselves to the 2D case, where the Hamiltonian is of the form  
\begin{eqnarray} H=\sum_{i=1}^2[{\textbf{p}_{i}^2\over 2 m^{*}}
+ {m^{*}\over 2}(\omega_{x}^2 x_{i}^2+\omega_{y}^2y_{i}^2)]
 +{{e^2}\over
\varepsilon^{*}|\textbf{r}_{2}-\textbf{r}_{1}|}.~~~~\label{ham}\end{eqnarray}
With $\varepsilon^{*}$ being the effective dielectric constant and $m^{*}$ the effective electron mass the above Hamiltonian is a frequently used model of the two-electron QD. 

After transformation
$\textbf{r}\mapsto \sqrt{2\hbar\over {{m^{*}\omega_{x}}}}\textbf{r}$,  $E\mapsto {\hbar \omega_{x}E\over {2}}$,
the Schr\"{o}dinger equation takes a form
\be H\Psi(\textbf{r}_{1},\textbf{r}_{2})=E\Psi(\textbf{r}_{1},\textbf{r}_{2}),\label{EOM} \ee
where the Hamiltonian is given by
\begin{eqnarray} H= \sum_{i=1}^2[-{1\over
2}\triangle_{\textbf{r}_{i}}+2
x_{i}^2+2\epsilon^2 y_{i}^2]
 +{g\over
|\textbf{r}_{2}-\textbf{r}_{1}|}
\label{hamrt}.\end{eqnarray} The dimensionless coupling $g={e^2\over\varepsilon^{*}}\sqrt{{2m^{*}\over \omega_{x}\hbar^{3}}}$
represents the ratio of the Coulomb repulsion to the confinement energy and the dimensionless parameter $\epsilon={\omega_{y}\over \omega_{x}}$ measures the anisotropy of the confining potential.

We will analyze the dependence of the entanglement between the particles in the ground-state of the system on the interaction strength $g$ and the anisotropy parameter $\epsilon$, paying particular attention to the regime of large $g$. In the case of finite $\epsilon$, the limit of $g\rightarrow \infty$
corresponds to the situation in which both frequencies of the trap tend to zero. In this case, regardless of the value of $\epsilon$, the correlations play an essential role. In this paper we provide a method for determining the natural orbitals in the $g\rightarrow \infty$ limit by applying the harmonic approximation to the anisotropic confinement case. We derive an explicit representation of the asymptotic natural orbitals in terms of one-dimensional orbitals defined by integral equations. This enables easy determination of the asymptotic occupancies and entanglement entropies for the whole range of $\epsilon$. For finite values of $g$ we determine the numerically exact results with the Rayleigh-Ritz method and demonstrate how the asymptotic values are attained. For all anisotropies, including the isotropic limit $\epsilon=1$, the asymptotic values of entanglement entropies are properly determined by the harmonic approximation.

The paper is arranged as follows. In section \ref{sec2} we discuss the two-particle state characteristics. In section \ref{3} we show the reliability of the harmonic approximation in the regime of large $g$ and provide the  asymptotic Slater-Schmidt decomposition. In this section the entanglement properties are examined in detail.  Finally, in section \ref{sumary} we make the concluding remarks.

\section{Two-particle state characteristics}\label{sec2}
\subsection{Energy eigenspectrum}
Consider a 2D system consisting of two identical fermions with a Hamiltonian given by (\ref{hamrt}).
 Since the Hamiltonian  does not depend on spin, the solution of the Schr\"{o}dinger equation
\be H\Psi(\zeta_{1},\zeta_{2})=E\Psi(\zeta_{1},\zeta_{2}),\label{Sch}\ee where
$\zeta_{i}=(\textbf{r}_{i},\sigma_{i})$, factorizes to the form
\be \Psi^{S\atop T}(\zeta_{1},\zeta_{2})=\chi^{\mp}_{s_{z}}\psi^{\pm}(\textbf{r}_{1},\textbf{r}_{2})
,\label{stste}\ee
where $s_{z}=\sigma_{1}+\sigma_{2}$ and the labels correspond to the singlet (S) and triplet (T) states, the spin functions of which are given by
$\chi^{\mp}_{s_{z}=0}= {\frac{1}{\sqrt{2}}}({\scriptstyle|\frac{1}{2}>}{\scriptstyle |-\frac{1}{2}>} \mp{\scriptstyle|-\frac{1}{2}>}{\scriptstyle |\frac{1}{2}>})$ and
$\chi^{+}_{s_{z}=\pm 1}= {\scriptstyle|\pm\frac{1}{2}>}{\scriptstyle |\pm\frac{1}{2}>}$. The spatial wavefunctions $\psi^{\pm}$, that are symmetric $(+)$ or  antisymmetric $(-)$
under permutation of the electrons, may be chosen real, since the interaction and confinement potentials are real functions.

Introducing the center of mass $\textbf{R}={1\over
2}(\textbf{r}_{1}+\textbf{r}_{2})=(X,Y)$
and relative  coordinates $\textbf{r}=\textbf{r}_{2}-\textbf{r}_{1}=(x,y)$, the Hamiltonian (\ref{hamrt}) may be written as $H=H^{\textbf{R}}+H^{\textbf{r}}$, where
\be H^{\textbf{R}}={-\nabla_{R}^2/
4}+4( X^2+\epsilon^2Y^2 ),\ee
 \be H^{\textbf{r}}=-\nabla_{r}^2 +
x^2+\epsilon^2 y^2 +{g\over r}\label{relham}.\ee
With wavefunction represented as a product $\Psi(x,y,X,Y)=\psi^{R}(X,Y)\psi^{r}(x,y),$ the
Schr\"{o}dinger equation (\ref{EOM}) separates into two equations
\be H^{\textbf{R}}\psi^{R}(\textbf{R})=E^{R}\psi^{R}(\textbf{R}),\label{cmrel}\ee
\be H^{\textbf{r}}\psi^{r}(\textbf{r})=E^{r}\psi^{r}(\textbf{r}),\label{rel}\ee
where the total energy $E=E^{r}+E^{R}$. As the center-of-mass coordinate remains the same upon the interchange of electrons, the symmetry requirement reduces to the symmetry of the relative wavefunction under inversion $\mathbf{r}\rightarrow
-\mathbf{r}$. Because of the invariance of $H^{\textbf{r}}$ to reflections about the $x-$ and $y-$axes, the $(x, y)-$
parity of $\psi^{r}(x,y)$ is well defined. The parity $(+, +)$ or $(-,-)$ corresponds to spin singlet eigenfunctions, and the parity $(+,-)$ or
$(-, +)$ to the spin triplet ones. The solutions of the CM equation (\ref{cmrel}) are given by
\be \psi^{R}_{nm}(X,Y)=\varphi_{n}(2X)\varphi_{m}(2\sqrt{\epsilon}Y),\ee
with the HO eigenfunction
\begin{eqnarray}\varphi_{n}(x)= e^{-x^2/2}
H_{n}(x),\label{cic}\end{eqnarray}
and corresponding energies $E^{R}_{nm}=2(n+{1\over
2})+2\epsilon(m+{1\over 2})$.
The relative-motion equation (\ref{rel}) is separable only if $\epsilon=1,2,1/2$. The first
case, $\omega_{x}=\omega_{y}$, being separable in polar coordinates, was much studied in the literature \cite{Merkt,Zhu1,wkb} and
in addition the closed-form solutions for particular frequencies have been
derived \cite{tautB}. In the case $\omega_{y}=2\omega_{x}$, Eq.(\ref{rel}) is separable in parabolic coordinates \cite{anis} and
the closed-form solutions may be also obtained for particular values of $\omega_{x}$~\cite{kos}. For other ratios of confinement frequencies the 2D Schr\"{o}dinger equation has to be solved by numerical techniques.
\subsection{Entanglement measures}\label{Entanglement}
Entanglement is a term used to describe quantum correlations between the particles. A convenient tool to analyze those correlations is the reduced density matrix (RDM) defined as \cite{redu}
\be \rho_{red}(\zeta,\zeta^{'})= Tr_{\zeta_{2}}(\langle\zeta,\zeta_{2}|\Psi\rangle\langle \Psi|\zeta_{2},\zeta^{'}\rangle).\label{RDM}\ee
For identical fermions the bi-partite pure state $|\Psi\rangle$ can be expressed as a combination of the Slater determinants made out of one-particle spin-orbitals in which its RDM (\ref{RDM}) is diagonal \cite{slater2,Lew, Ghirardi}. The number of expansion coefficients appearing in the Slater decomposition, that are different from zero, is called the Slater rank ($SR$). The pure state of identical particles is considered nonentangled if $SR=1$, i.e. the only correlations that exist between the fermions can be attributed to their indistinguishable nature \cite{Ghirardi}.

The total wavefunction factorizes into spatial and spin components  (\ref{stste}) and the same holds for the reduced density matrix (\ref{RDM})
\be \rho_{red,s_{z}}^{\pm}(\zeta,\zeta^{'})=\rho_{red,s_{z}}^{\mp}(\sigma,\sigma{'})
\rho_{red}^{\pm}(\textbf{r},\textbf{r}^{'}),\ee
where
$\rho_{red}^{\pm}(\textbf{r},\textbf{r}^{'})=\int\psi^{\pm}(\textbf{r},\textbf{r}_{2})\psi^{\pm}(\textbf{r}_{2},\textbf{r}^{'})
d\textbf{r}_{2}$
and the spin parts
$\rho^{-}_{red,s_{z}=0}=\rho^{+}_{red,s_{z}=0}=diag(1/2 ,1/2)$,   $\rho^{+}_{red,s_{z}=1}=diag(1,0)$ and $\rho^{+}_{red,s_{z}=-1}=diag(0,1)$.

The eigenvalue problem for the spatial part $\rho^{\pm}_{red}(\textbf{r},\textbf{r}^{'})$
can be written in the form
\be \int\rho_{red}^{\pm}(\textbf{r},\textbf{r}^{'})v_{l}^{\pm}(\textbf{r}^{'})
d\textbf{r}^{'}=\lambda_{l}^{\pm}v_{l}^{\pm}(\textbf{r}),\label{rrd}\ee
which determines the natural orbitals $v_{l}^{\pm}$ and the occupancies $\lambda_{l}^{\pm}$. The families  $\{v_{l}^{+}(\textbf{r})\}$ and $\{v_{l}^{-}(\textbf{r})\}$ form orthonormal basis sets in the space of symmetric and antisymmetric functions, respectively. The spatial parts of two-particle functions may be represented in terms of natural orbitals. Representation for the symmetric function takes the form of the Schmidt decomposition \be \psi^{+}(\textbf{r}_{1},\textbf{r}_{2})=\sum_{l} k_{l}^{+} v_{l}^{+}(\textbf{r}_{1}) v_{l}^{+}(\textbf{r}_{2}),\label{rrs}\ee
and in the case of antisymmetric function the Slater decomposition holds \cite{Ghirardi}
\be \psi^{-}(\textbf{r}_{1},\textbf{r}_{2})=\sum_{l} {k_{l}^{-}\over \sqrt{2}}[ v_{2l-1}^{-}(\textbf{r}_{1}) v_{2l}^{-}(\textbf{r}_{2})-v_{2l-1}^{-}(\textbf{r}_{2}) v_{2l}^{-}(\textbf{r}_{1})].\label{rra}\ee
The coefficients $k_{l}^{+}$ and $k_{l}^{-}$
are related to the eigenvalues of (\ref{rrd})  by $\lambda_{l}^{+}=[k_{l}^{+}]^2$ and  $\lambda_{l}^{-}={[k_{l}^{-}]^2\over 2}$, respectively. The eigenvectors  $v_{2l}^{-}$ and $v_{2l-1}^{-}$ correspond to the same eigenvalue $\lambda_{l}^{-}$, which means that eigenvalues of the spatial RDM of antisymmetric wavefunction are doubly degenerate and satisfy the conservation of probability $2\sum_{l}\lambda_{l}^{-}=1$. We assume that the eigenvalues $\lambda_{l}^{\pm}$ are ordered such that $\lambda_{0}^{\pm}\geq \lambda_{1}^{\pm} \geq...$. It can be easily inferred from the appropriate decompositions of the spatial parts (\ref{rrs}) or (\ref{rra}) that the $SR$ is related to the number $n$ of non-vanishing eigenvalues of the spatial RDM (being not necessarily different) as follows: the singlet and the triplet states with  $s_{z}=0$ have $SR=n$, whereas the triplet states with $s_{z}=-1,1$ have $SR=n/2$. The two-particle state is non-entangled if and only if $SR=1$ and deviations from such a form may be used to measure the amount of entanglement in the system.

The entanglement depends on the whole spectrum of the RDM and its characteristics may be constructed from the natural orbitals occupancies $\lambda_{l}$ \cite{vgh2,karol,vgh1}. In this work we consider the von Neumann (vN) entropy and the linear entropy  that are the most popular measures of entanglement in pure states. The vN entropy
\cite{vn1,vn2,Ghirardi} is defined as
\be  \textbf{S}=-Tr[{\rho}_{red} Log_{2}{\rho}_{red}]\label{vnnn},\ee
and in the two-particle case separates into
\begin{eqnarray}
\textbf{S}=S^{spin}+S^{vN},\end{eqnarray}
where the spin contribution $S^{spin}=1$ if $s_{z}=0$ and $S^{spin}=0$ if $s_{z}=\pm 1$.

The space part that depends on interactions may be calculated with the RDM eigenvalues as
\begin{eqnarray}
S^{vN}=-\sum_{l=0}\lambda_{l} Log_{2}\lambda_{l}.\end{eqnarray}

The linear entropy \cite{le}
 \be \textbf{L}=1-Tr \rho_{red}^2\label{lin}\ee
can be calculated directly from the above definition, which is a big advantage over the case of vN entropy (\ref{vnnn}) since diagonalization of the RDM is not needed.

\section{Harmonic approximation}\label{3}

 \subsection{Approximate spectrum}\label{ts}
The harmonic
approximation was applied with success in the case of circular \cite{tautB,MatulisPeeters,Puente} and anisotropic confinement potential \cite{anis19}.
The anisotropic potential $V(x,y)= x^2
+\epsilon^2 y^2 +{g/ {\sqrt{{x}^2+y^2}}}$  has two local minima at  $(x_{cl},0)$ and $(-x_{cl},0)$, where $x_{cl}=(g/2)^{1\over 3}$ is the classical equilibrium distance
between the Coulombically interacting particles in the trap.
Expanding $V(x,y)$ into a Taylor series around the minimum at $\textbf{r}_{min}=(x_{cl},0)$ and retaining the terms up to second order, the relative
motion equation (\ref{rel}) gets approximated by the
Schr\"{o}dinger equation \begin{eqnarray}[-\bigtriangleup_{r} + V(\textbf{r}_{min})+3
(x-x_{cl})^2+(\epsilon^{2}-1)y^2]\psi^{r}\nonumber\\=E^{r}\psi^{r}
,\label{Schap}\end{eqnarray}
the eigenvalues of which
are of the form \be E^{r}_{nm}=V(\textbf{r}_{min})+2\sqrt{3}({1\over
2}+n)+2({1\over
2}+m)\sqrt{\epsilon^{2}-1}.\label{aprox}\ee The
corresponding eigenfunctions

\be\psi_{nm}^{r}(x,y)=\varphi_{n}(3^{1/4}(x-x_{cl}))\varphi_{m}((\epsilon^2-1)^{1/4}y),\label{wf}\ee
where $\varphi_{n}$ is given by (\ref{cic}), provide approximations to the relative motion wavefunctions only around $(x_{cl},0)$. The approximations around  $(-x_{cl},0)$ are obtained by the transformation $x\mapsto-x$. The states of relative motion with well-defined parity $(x,y)$ may be constructed as
\be
\psi^{r, \pm}_{nm}(x,y)=\psi^{r}_{nm}(x,y)\pm\psi^{r}_{nm}(-x,y),\label{apfun1}\ee
where the sign $+$/$-$ relates to the even/odd $x-$parity. Notice that each energy level of the relative motion in the
harmonic  approximation is doubly degenerate,
namely the states with spatial $x$-$y$
parity:  $(-, +)$,  $(+,-)$ are degenerate with those of
$(+, +)$, $(-,-)$, respectively.

\begin{figure}[h]
\begin{center}
\includegraphics[width=0.475\textwidth]{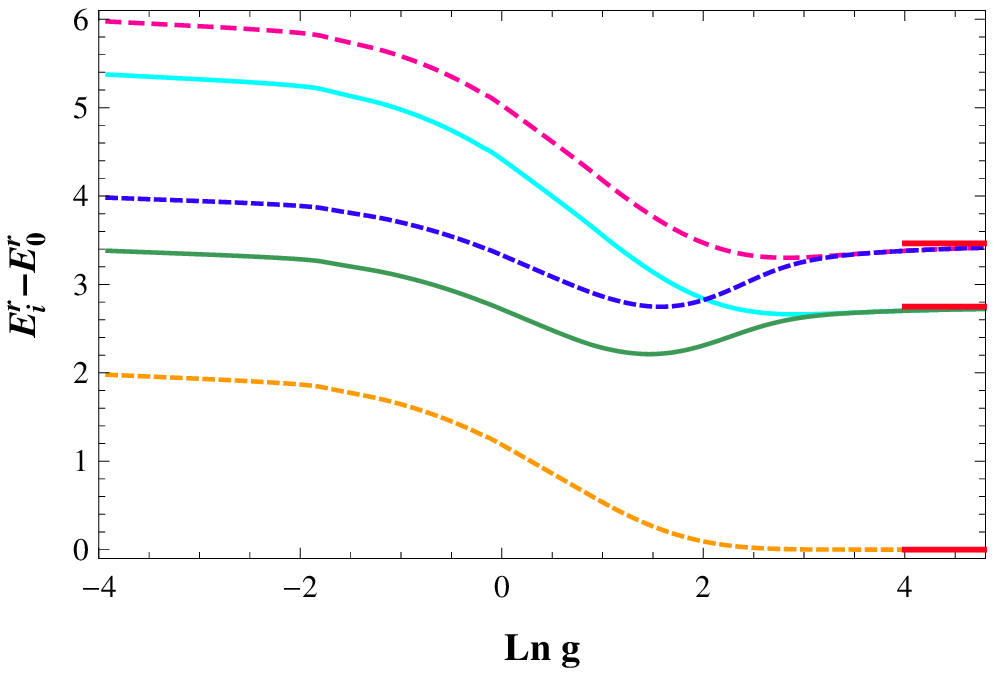}
\includegraphics[width=0.475\textwidth]{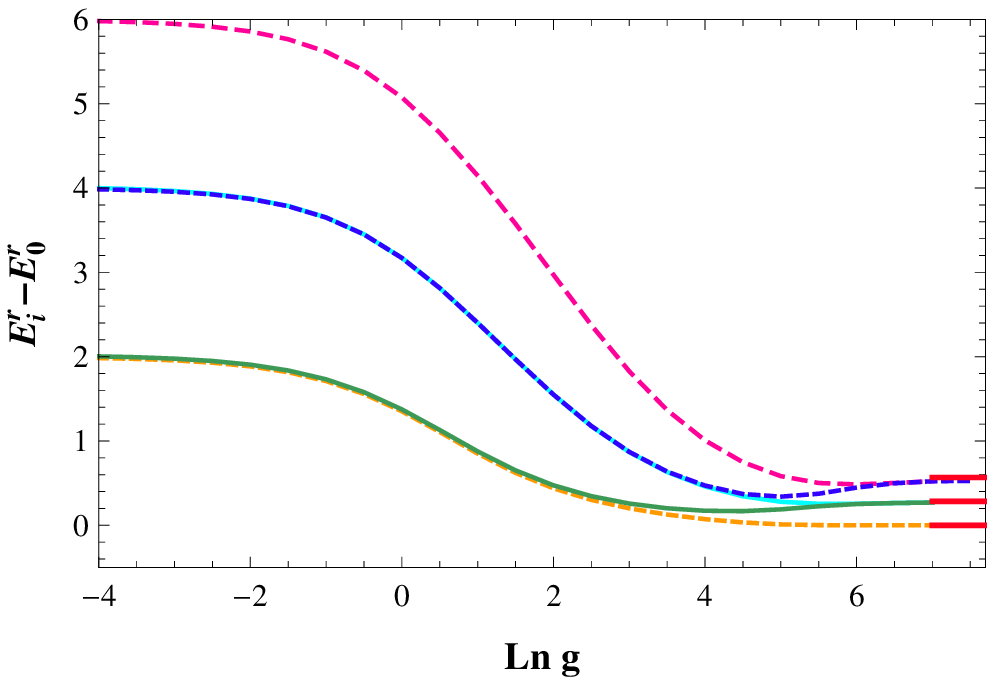}
\end{center}
 \caption{\label{fig25ee3f:beh} Excitation energies $E^{\textbf{r}}_{i}-E^{\textbf{r}}_{0}$ in function of $\ln g$ for $\epsilon=1.7$ (top) and $\epsilon=1.01$ (bottom). The horizontal lines mark the asymptotic results of the harmonic approximation.}\end{figure}

The quality of harmonic approximation has been tested by comparison with  numerical solutions of the relative motion equation (\ref{rel}) obtained
through the exact diagonalization method in the basis of the two-dimensional harmonic oscillator eigenfunctions of the required parity. In Fig. \ref{fig25ee3f:beh} the excitation energies $E^{r}_{i}-E^{r}_{0}$ of the Hamiltonian (\ref{relham}) are plotted in function of $\ln g$ for a large ($\epsilon=1.7$) and a small ($\epsilon=1.01$) anisotropy. The horizontal lines mark the results of the harmonic approximation (\ref{aprox}), which becomes exact at $g\rightarrow\infty$. We observe that for smaller values of $\epsilon$ the asymptotic behavior of the relative motion energies is reached at larger values of $g$. This is consistent with the fact that the closer the value of $\epsilon$ to one, the larger the value of $g$ at which the harmonic approximation is applicable.

 For transparency of presentation we concentrate on analyzing the singlet and triplet states of the lowest energy that become degenerate in the limit of $g\rightarrow \infty$. They correspond to the ground state of CM; thus the total spatial wavefunctions are $\psi^{+}=\psi^{r,+}_{00}\psi^{R}_{00}$ and $\psi^{-}=\psi^{r,-}_{00}\psi^{R}_{00}$, respectively. Harmonic approximations to those functions can be written in the form convenient for further analysis
\begin{eqnarray}\psi^{\pm}(x_{1},x_{2},y_{1},y_{2})=C_{\pm}(g,\epsilon)
h(y_{1},y_{2})(q(x_{1},x_{2})\pm q(x_{2},x_{1})),\label{foplk}\end{eqnarray}
where
\begin{eqnarray} q(x_{1},x_{2})=\varphi_{0}(3^{1/4}(x_{2}-x_{1}-x_{cl}))\varphi_{0}(x_{2}+x_{1})= \nonumber\\e^{-{1\over 2}[{\sqrt{3}}(x_{2}-x_{1}-x_{cl})^2+(x_{1}+x_{2})^2]},\label{gg}\end{eqnarray}
\begin{eqnarray} h(y_{1},y_{2})=\varphi_{0}((\epsilon^2-1)^{1/4}(y_{2}-y_{1}))
\varphi_{0}(\sqrt{\epsilon}({y_{2}+y_{1}}))= \nonumber\\e^{-{1\over 2}[\sqrt{\epsilon^2-1}(y_{1}-y_{2})^2+\epsilon(y_{1}+y_{2})^2]},\end{eqnarray}
and the normalization constants are given by
\be C_{\pm}(g,\epsilon)= {\sqrt{2}3^{{1\over 8}}\epsilon^{{1\over 4}}(\epsilon^2-1)^{{1\over 8}}\over \pi \sqrt{1\pm e^{-{\sqrt{3} {({g\over 2})^{2\over 3}} }}}}\overrightarrow{~g\rightarrow \infty~}{\sqrt{2}3^{{1\over 8}}\epsilon^{{1\over 4}}(\epsilon^2-1)^{{1\over 8}}\over \pi}.\ee

\subsection{The linear entropy}

Quantum entanglement in two-particle systems is usually quantified by entropic measures. The easiest to calculate is the linear entropy that may be obtained directly from the definition (\ref{lin}) using the RDM calculated from the spatial wavefunction. The asymptotic expression for wavefunction (\ref{foplk}) enables us to determine the linear entropy in the limit of $g\rightarrow\infty$. For the lowest singlet state the explicit expression reads
\begin{eqnarray}
\textbf{L}^{ g\rightarrow \infty}(\epsilon)=1- {3^{1\over 4}(\epsilon^2-1)^{1\over 4}\sqrt{\epsilon(1-{\sqrt{3}\over 2})}\over \epsilon+\sqrt{\epsilon^2-1}}.\label{liny}\end{eqnarray}
Although the above formula is strictly valid only if $\epsilon>1$, since it has been derived from the harmonic approximation, we observe that its limit at $\epsilon\rightarrow1^{+}$ is equal to one and coincides thus with the exact asymptotic value of the linear entropy.

\begin{figure}[h]
\begin{center}
\includegraphics[width=0.50\textwidth]{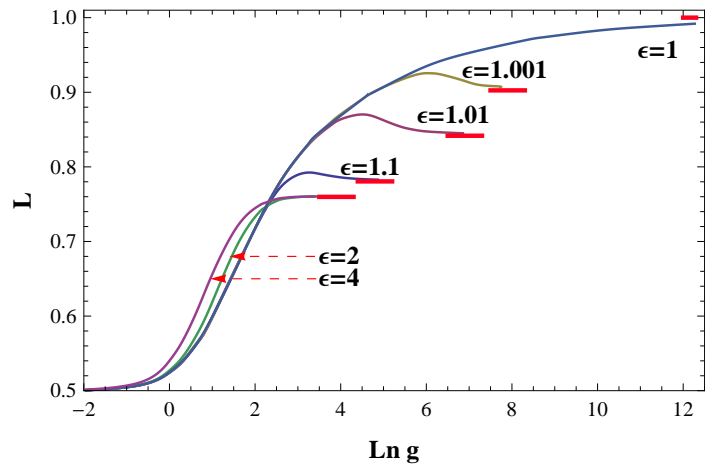}
\end{center}
 \caption{\label{fig2ff65e3f:beh}The linear entropy of the lowest singlet state for various anisotropy ratios $\epsilon$ in function of $\ln g$. The horizontal lines show the asymptotic values at $g\rightarrow \infty$. }
 \end{figure}

In Fig.\ref{fig2ff65e3f:beh} we show how the linear entropy approaches the asymptotic limits for various values of the anisotropy parameter. The results at finite values of $g$ have been obtained by numerical integration with the use of wavefunctions determined by the Rayleigh-Ritz procedure. At large anisotropy the dependence on $g$ is very similar and the asymptotic values of the linear entropy for $\epsilon=2$ and $\epsilon=4$ are nearly equal to that in the infinite anisotropy limit $\textbf{L}^{ g\rightarrow \infty}(\epsilon\rightarrow \infty)\approx 0.759142$. Only at $\epsilon\lesssim 1.4$ asymptotic values of the linear entropy start to
visibly differ from $0.759142$
and reach the value one at  $\epsilon \rightarrow1^{+}$.

In this range of $\epsilon$ the linear entropy exhibits a local maximum and approaches monotonically from above its asymptotic value determined by the formula (\ref{liny}), which confirms its validity. The smaller is the anisotropy, the larger is the value of $g$ at which it occurs, which is in accordance with our earlier discussion. The dependence of the linear entropy on $\epsilon$ becomes strong for large enough $g$, but only in a small vicinity of $\epsilon=1$.
\subsection{Asymptotic Slater-Schmidt decomposition}\label{smidt}
\begin{figure}[h]
\begin{center}
\includegraphics[width=0.50\textwidth]{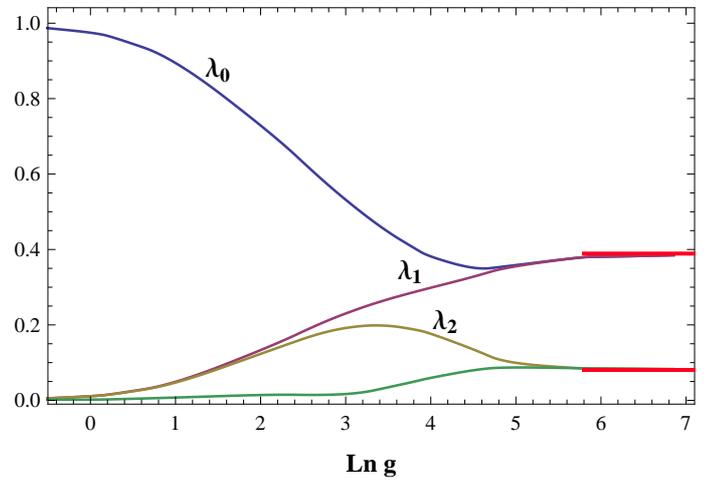}
\end{center}
 \caption{\label{occup} Four largest occupancies in function of $\ln g$ at fixed anisotropy $\epsilon=1.01$.  The asymptotic values are marked by horizontal lines.} \end{figure}

\begin{figure}[h]
\begin{center}
\includegraphics[width=0.50\textwidth]{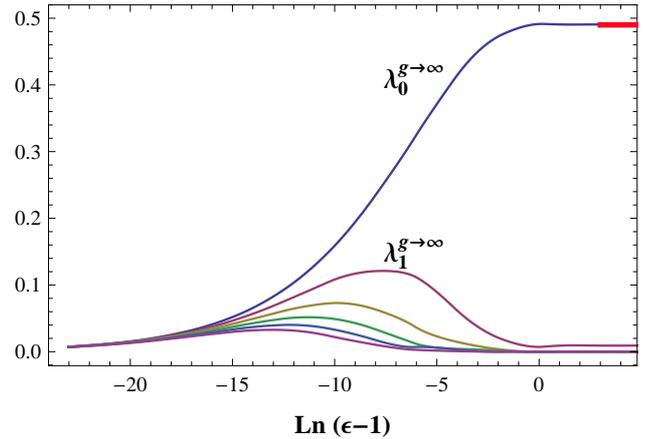}
\end{center}
 \caption{\label{fig25e3f:beh} Behavior of the six largest asymptotic occupancies in function  of  $\ln(\epsilon-1)$. The  limit  of the largest occupancy $\lambda _{0}^{g\rightarrow\infty}\approx 0.490688$  as $\epsilon\rightarrow\infty$ is marked by a horizontal line.}\end{figure}

In order to calculate the vN entropy, determination of the natural orbital occupancies is necessary. The asymptotic behavior at $g\rightarrow\infty$ may be determined by deriving the Schmidt decomposition of the asymptotic two-particle function (\ref{foplk}). To this end it is convenient to introduce new coordinates
$x_{1}\mapsto{\tilde{x}_{1}}-{x_{cl}\over 2},x_{2}\mapsto\tilde{x}_{2}+{x_{cl}\over 2},$
so as the function $q$ (\ref{gg}) transforms into \be q(x_{1},x_{2})\mapsto \tilde{q}(\tilde{x}_{1},\tilde{x}_{2})=e^{-{1\over 2}[\sqrt{3}(\tilde{x}_{2}-\tilde{x}_{1})^2+(\tilde{x}_{1}+\tilde{x}_{2})^2]}.
\label{hjk}\ee
Being real and symmetric, the function $\tilde{q}$ has the Schmidt decomposition
\be \tilde{q}(\tilde{x}_{1},\tilde{x}_{2})=\sum_{n=0}k_{n}^{(1)}\vartheta_{n}^{(1)}(\tilde{x}_{1})
\vartheta_{n}^{(1)}(\tilde{x}_{2}).\label{s1}\ee
It is worthwhile to notice that the function (\ref{hjk}) is $g$-independent and so are
the orbitals $\vartheta_{n}^{(1)}$ and the corresponding expansion coefficients $k_{n}^{(1)}$. They can be determined from the integral equation
 \be \int_{-\infty}^{\infty}\tilde{q}(x,x^{'})\vartheta_{n}^{(1)}(x^{'})dx^{'}=
 k_{n}^{(1)}\vartheta_{n}^{(1)}(x),\label{eq11} \ee
Changing the variables back to $x_{1}$ and $x_{2}$, one gets
\be q(x_{1},x_{2})=\sum_{n=0}k_{n}^{(1)}\vartheta_{n}^{(1)}(x_{1}+{x_{cl}\over 2})\vartheta_{n}^{(1)}(x_{2}-{x_{cl}\over 2}).\label{s1}\ee
The function $h(y_{1},y_{2})$ is real and symmetric, thus its Schmidt decomposition reads
\be h(y_{1},y_{2})=\sum_{m=0}k_{m}^{(2)}\vartheta_{m}^{(2)}(y_{1})\vartheta_{m}^{(2)}(y_{2}),\label{s2}\ee
where $\vartheta_{m}^{(2)}$ and $k_{m}^{(2)}$ are determined by
\be \int_{-\infty}^{\infty}h(y,y^{'})\vartheta_{m}^{(2)}(y^{'})dy^{'} =k_{m}^{(2)}\vartheta_{m}^{(2)}(y),\label{eq2} \ee
with the orthogonal orbitals $\vartheta_{m}^{(2)}$ assumed to be normalized to unity. Using the expansions (\ref{s1}) and (\ref{s2}) we represent the wavefunctions (\ref{foplk}) as
\be\psi^{\pm}(\textbf{r}_{1},\textbf{r}_{2})=\sum_{l=0} k_{l}^{\pm}[L_{l}(\textbf{r}_{1})R_{l}(\textbf{r}_{2})\pm R_{l}(\textbf{r}_{1})L_{l}(\textbf{r}_{2})],\label{edee}\ee
where $l=(n,m)$,  $k_{l}^{\pm}=C_{\pm}(g,\epsilon) k_{n}^{(1)}k_{m}^{(2)}$ and 
\begin{eqnarray}L_{l}(\textbf{r})=\vartheta_{l}(x+{x_{cl}\over 2},y)=\vartheta_{n}^{(1)}(x+{x_{cl}\over 2})\vartheta_{m}^{(2)}(y),\nonumber\\ R_{l}(\textbf{r})=\vartheta_{l}(x-{x_{cl}\over 2},y)=\vartheta_{n}^{(1)}(x-{x_{cl}\over 2})\vartheta_{m}^{(2)}(y), \nonumber\end{eqnarray}
are the one-particle orbitals centered around classical  equilibrium points which satisfy $\langle L_{l}|L_{k}\rangle=\delta_{kl}$ and
 $\langle R_{l}|R_{k}\rangle=\delta_{kl}$. If we define the
 new orbitals
 \begin{eqnarray}v_{l}(\textbf{r})={R_{l}(\textbf{r})+L_{l}(\textbf{r})\over\sqrt{2}},u_{l}(\textbf{r})={L_{l}
(\textbf{r})-R_{l}(\textbf{r})\over\sqrt{2}},\label{cons}\nonumber\end{eqnarray}
that fulfil $\langle u_{l}|v_{l}\rangle=0$, the spatial wavefunctions can be expressed as
\be \psi^{+}(\textbf{r}_{1},\textbf{r}_{2})=\sum_{l=0} k_{l}^{+}[v_{l}(\textbf{r}_{1})v_{l}(\textbf{r}_{2})-u_{l}
(\textbf{r}_{1})u_{l}(\textbf{r}_{2})],\label{ed}\ee
and
\be \psi^{-}(\textbf{r}_{1},\textbf{r}_{2})=\sum_{l=0} k_{l}^{-}[u_{l}(\textbf{r}_{1})v_{l}(\textbf{r}_{2})-v_{l}
(\textbf{r}_{1})u_{l}(\textbf{r}_{2})].\label{ed1}\ee
In the limit of $g\rightarrow\infty$ we have $k_{l}^{+}=k_{l}^{-}=k_{l}$ and the integral overlap $\langle L_{l}|R_{k} \rangle $ vanishes for any $l,k$, and because of that $||\psi^{\pm}||^2=2\sum_{l}k_{l}^2$,
$\langle u_{l}|u_{k}\rangle=\langle v_{l}|v_{k}\rangle=\delta_{lk}$, $\langle u_{l}|v_{k}\rangle=0$. In this limit the symmetric spatial wavefunction coincides with the absolute value of the antisymmetric one  $ \psi^{+}(\textbf{r}_{1},\textbf{r}_{2})=|\psi^{-}(\textbf{r}_{1},
\textbf{r}_{2})|$.  Moreover, (\ref{ed}) and (\ref{ed1}) yield the  Schmidt decomposition (\ref{rrs}) and  the Slater decomposition (\ref{rra}), respectively. In both cases the decomposition of the spatial RDM reads
   \be \rho^{\pm}_{red}(\textbf{r},\textbf{r}^{'})=\sum_{l=0}k_{l}^2[v_{l}(\textbf{r})v_{l}(\textbf{r}^{'})+u_{l}(\textbf{r})u_{l}(\textbf{r}^{'})],\label{ei}\ee
which shows that the eigenvectors $v_{l}$ and $u_{l}$ correspond to the same occupancy $\lambda_{l}={2 3^{{1\over 4}}\epsilon^{{1\over 2}}(\epsilon^2-1)^{{1\over 4}}\over \pi^2}( k_{n}^{(1)}k_{m}^{(2)})^{2}$ with the coefficients $k_{n}^{(1)}$ and $k_{n}^{(2)}$ determined by Eqs. (\ref{eq11}) and (\ref{eq2}), respectively. The above equations can easily be solved  through discretization technique.
By discretizing the $x$($y$) and $x^{'}$($y^{'}$) variables with equal subintervals of length $\triangle x$ ($\triangle y$), the integral equations turn into algebraic eigenvalue problems
\be \sum_{s}[A_{rs}^{(i)}-\delta_{rs}{k_{n}^{(i)}}]{\vartheta_{n}^{(i)}}
(z_{s})=0,~r,i=1,2,\label{ppp1}\ee
where $A_{rs}^{(1)}=\tilde{q}(x_{r},x_{s})\triangle x $ and
$A_{rs}^{(2)}= h(y_{r},y_{s})\triangle y$.
Diagonalization of the matrix $[A_{rs}^{(i)}]_{N\times N}$ provides thus a set of approximations to $N$  of the largest modulus eigenvalues $k_{n}^{(i)}$. One has to notice that the function $h$ depends on $\epsilon$ and the range of argument where its value is non-negligible strongly increases when $\epsilon\rightarrow 1^{+}$. Therefore, in the case of small anisotropy, an appropriately large interval must be discretized to achieve a reasonable accuracy in the diagonalization of $[A_{rs}^{(2)}]_{N\times N}$. In Fig. \ref{occup} the four largest natural orbital occupancies determined in function of $g$ by the numerically exact Rayleigh-Ritz procedure are shown for the nearly isotropic example of $\epsilon=1.01$. It is seen how they converge into asymptotic doublets of the values determined in the harmonic approximation by the procedure described above.
The behavior of the asymptotic occupancies at varying anisotropy ratio  $\epsilon$ is shown in Fig. \ref{fig25e3f:beh}. For a better display the results are presented in function of $\ln(\epsilon-1)$, since the occupancies are strongly sensitive to small changes in $\epsilon$ around $\epsilon= 1$. In the limit of $\ln(\epsilon-1)\rightarrow-\infty$, the values of all asymptotic occupancies tend to zero in such a way that their sum is equal to $\frac{1}{2}$. Notice that this behavior is related to the degeneracy that appears in the energy spectrum of the system in the circular symmetry limit. The clustering of occupancies around zero value is clearly visible in Fig.\ref{fig25e3f:beh}. With increasing anisotropy the situation changes dramatically, namely a tiny deviation from circular symmetry leads to the lifting of the clustering. The value of $\ln(\epsilon-1)\approx -10$, at which the effect is clearly visible,
corresponds to a very small anisotropy, $\epsilon\approx 1.00005$.
At higher confinement anisotropies,
all the occupancies but the largest one exhibit a local maximum. The larger the value of the occupancy, the larger is the value of $\epsilon$ at which the maximum occurs. Above the critical threshold $\epsilon_{cr}\approx 1.4$ ($\ln(\epsilon_{cr}-1)\approx -1$) all those occupancies saturate at  vanishingly small values. The largest occupancy $\lambda_{0}^{g\rightarrow\infty}$ performs differently, as it grows monotonically with increasing $\epsilon$ at the cost of the remaining occupancies and saturates above $\epsilon_{cr}$.  The limit of $\ln(\epsilon-1)\rightarrow\infty$ corresponds to  $\epsilon\rightarrow \infty$, i.e. a  one dimensional motion along the $x$-axis. In this limit
$\lambda _{0}^{g\rightarrow\infty}\approx 0.490688$ and the sum of all the remaining occupancies $2\sum_{l=1}\lambda_{l}^{g\rightarrow \infty}$ is only about $0.018624$, which means that the natural orbitals  $v_{0}(\textbf{r})$ , $u_{0}(\textbf{r})$ are the only two that are substantially occupied. This indicates that in the case of strong interaction for enough anisotropic confinement ($\epsilon\gtrsim \epsilon_{cr}$), the terms higher than $l=0$ in (\ref{edee}) contribute very little and the spatial functions approach the form
\be\psi^{\pm}_{g\rightarrow\infty}(\textbf{r}_{1},\textbf{r}_{2})\approx{1\over \sqrt{2}}[L_{0}(\textbf{r}_{1})R_{0}(\textbf{r}_{2})\pm R_{0}(\textbf{r}_{1})L_{0}(\textbf{r}_{2})].\label{polo}\ee
The total singlet and triplet wavefunctions with $s_{z}=0$ are thus well approximated by
\begin{eqnarray}\Psi^{S\atop T}_{s_{z}=0}(\zeta_{1},\zeta_{2})\approx{\frac{1}{2}}({\scriptstyle |\frac{1}{2}>}{\scriptstyle |-\frac{1}{2}>} \mp{\scriptstyle |-\frac{1}{2}>}{\scriptstyle |\frac{1}{2}>})\times\nonumber\\( L_{0}(\textbf{r}_{1})R_{0}(\textbf{r}_{2})\pm R_{0}(\textbf{r}_{1})L_{0}(\textbf{r}_{2})).\label{polo0}\end{eqnarray}
Each of them constitutes a sum of two Slater determinants ($SR=2$) and represents an entangled state. The situation is different for the triplet components with $s_{z}= \pm 1$, since their total wavefunctions
\begin{eqnarray}\Psi^{T}_{s_{z}=\pm 1}(\zeta_{1},\zeta_{2})\approx\nonumber\\{\frac{1}{\sqrt{2}}}{\scriptstyle |\pm\frac{1}{2}>}{\scriptstyle |\pm\frac{1}{2}>} (L_{0}(\textbf{r}_{1})R_{0}(\textbf{r}_{2})- R_{0}(\textbf{r}_{1})L_{0}(\textbf{r}_{2}))\label{polo1}\end{eqnarray}
represent one Slater determinant ($SR=1$) and those states have to be regarded
as non-entangled. It has to be stressed however, that this is only an approximate result since even at $\epsilon\rightarrow \infty$ the other occupancies do not strictly vanish.

\subsection{Comparison of the linear and vN entropy}\label{smidt}
\begin{figure}[h]
\begin{center}
\includegraphics[width=0.5\textwidth]{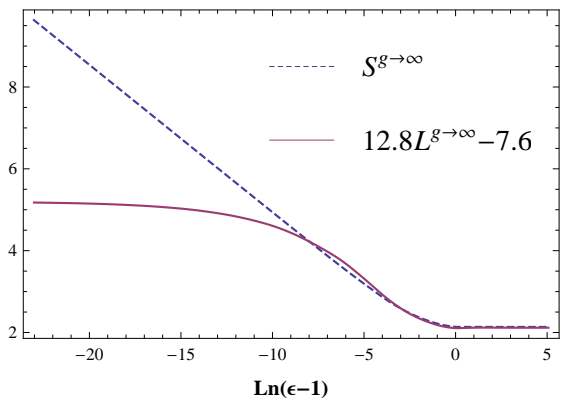}
\end{center}
 \caption{\label{fig265e3f:beh}Comparison of the asymptotic vN entropy $\textbf{S}^{g\rightarrow \infty}$ and the (appropriately rescaled) linear entropy $\textbf{L}^{g\rightarrow \infty}$ in function of $\ln(\epsilon-1)$.}
 \end{figure}

Since the linear entropy is relatively easy to calculate, several works proposed \cite{vgh1,vn2,linpro1,linpro3} to use it instead of the vN entropy arguing that their dependence on parameters of the system is similar. Here we test this conjecture for the lowest singlet state by comparing the behavior of both measures in function of $\epsilon$  at $g\rightarrow\infty$, where the differences between both measures are the most pronounced.
We observed that for $\epsilon$ large, the dependence between the vN and linear entropy is approximately linear and well fitted by $\textbf{S}^{g\rightarrow \infty}(\epsilon)\approx 12.8 \textbf{L}^{g\rightarrow \infty}(\epsilon)-7.6$.
The comparison of the so rescaled linear entropy with vN entropy is presented in Fig.
 \ref{fig265e3f:beh} in function of $\ln(\epsilon-1)$.
 Above $\epsilon_{cr}\approx 1.4$ both entropies saturate at constant values,  $\textbf{S}_{\epsilon\rightarrow\infty}^{g\rightarrow \infty}\approx 2.13618 $ and $\textbf{L}_{\epsilon\rightarrow\infty}^{g\rightarrow\infty}
 \approx 0.75914$, respectively. The closeness of those values to $2$ and $3/4$ can be attributed to the dominance of the two occupancies $\lambda_{0}^{+}\approx \lambda_{1}^{+}$ in this regime.
The vN entropy grows with decreasing $\epsilon$, which reflects the fact that the smaller is the value of $\epsilon$, the larger number of orbitals in the sum (\ref{edee}) becomes important.
The increase is nearly linear which means that the asymptotic vN entropy varies logarithmically with $\epsilon$
and tends to infinity when the confinement becomes circularly symmetric ($\epsilon\rightarrow 1^{+}$). Being a bounded function, the linear entropy performs differently at very small anisotropies, where its behavior starts to deviate from the linear one
and approaches the maximum possible value $1$ in the limit of $\epsilon\rightarrow 1^{+}$. Although in different ways, but both entanglement measures clearly demonstrate that the more circular is the confinement of strongly interacting particles, the more entangled is the system. The difference in behavior of the vN entropy and the linear entropy appears only in the vicinity of $\epsilon= 1$, where the system becomes degenerate.


\section{Summary}\label{sumary}
We performed a detailed examination of the entanglement properties for the system of two Columbically interacting electrons confined in a 2D anisotropic harmonic potential. The harmonic approximation has been developed in order to study the strongly interacting (weak confinement) case. Using the harmonic approximation we derived an explicit expression for the linear entropy of the lowest singlet state at $g\rightarrow\infty$. The occupancies $\lambda_{l}^{g\rightarrow\infty}$ in the asymptotically degenerate lowest singlet and triplet states may be easily determined numerically within this approximation. Performing numerical calculation through the Rayleigh-Ritz method we have also calculated the characteristics of the system at finite values of $g$. We demonstrated how the occupancies reach their asymptotic values as $g$ is increased. The asymptotic occupancies are strongly sensitive to changes of $\epsilon$  in the range of $1<\epsilon\lesssim \epsilon_{cr}\approx 1.4$, while above $\epsilon_{cr}$ they practically reach the values corresponding to infinite anisotropy $\epsilon\rightarrow \infty$, where the value of $\lambda_{0}^{g\rightarrow\infty}\approx 0.490688$ is the only substantial. This results in non-trivial entanglement properties since only two natural orbitals contribute significantly to the lowest singlet and triplet states. Furthermore, we have verified that the linear entropy is almost linearly related to the corresponding vN entropy except of the region of very small anisotropy $\epsilon\lesssim 1.00005$. Our calculations have shown that the entanglement is relatively insensitive to the shape of the harmonic confinement if the interaction is very weak (strong confinement case). In the strong correlation case the influence of anisotropy is much more pronounced. 

It would be interesting to carry out an analysis of entanglement for higher excited states. The thermal entanglement and the interaction with the environment will be the topic of further studies.

\bibliography{aipsamp}

\end{document}